%% file: energy-feedback.tex
\documentclass[9pt,conference]{IEEEtran}
\pdfoutput=1
\usepackage[utf8]{inputenc}
\usepackage[T1]{fontenc}
\usepackage{lmodern}  
\usepackage[british]{babel}
\usepackage{url}
\usepackage{gensymb} 

\def\WORKSHOP

\usepackage[table]{xcolor} 
\usepackage{colortbl,makecell,siunitx,booktabs,setspace,ctable,dcolumn}
\usepackage{pifont} 
\newcommand{\cmark}{\ding{51}}
\newcommand{\xmark}{\ding{55}}
\newcommand*\rot{\rotatebox{90}}

\newcolumntype{C}[1]{>{\setlength{\baselineskip}{1em}\centering\let\newline\\\arraybackslash\hspace{0pt}}m{#1}}

\newcolumntype{P}[1]{>{\setlength{\baselineskip}{1em}\centering\let\newline\\\arraybackslash\hspace{0pt}}p{#1}}

\newcolumntype{R}[1]{>{\raggedleft\let\newline\\\arraybackslash\hspace{0pt}}m{#1}}
\newcolumntype{d}[1]{D{.}{\cdot}{#1}}
\newcommand{\dmake}[3]{\makecell[{{S[table-format=#1.#2]}}]{#3}}

\usepackage[
  backend=bibtex,
  style=ieee, 
  doi=false,
  isbn=false,
  url=false,
  maxbibnames=99]{biblatex}
\DeclareFieldFormat{eprint:citeseerx}{%
  \small{CiteSeerX DOI}\addcolon\space
  \ifhyperref
    {\href{http://citeseerx.ist.psu.edu/viewdoc/summary?doi=#1}{\nolinkurl{#1}}}
    {\nolinkurl{#1}}}

\newbibmacro{string+doiurlisbn}[1]{%
  \iffieldundef{doi}{%
    \iffieldundef{url}{%
      \iffieldundef{isbn}{%
        \iffieldundef{issn}{%
          #1%
        }{%
          \href{http://books.google.com/books?vid=ISSN\thefield{issn}}{#1}%
        }%
      }{%
        \href{http://books.google.com/books?vid=ISBN\thefield{isbn}}{#1}%
      }%
    }{%
      \href{\thefield{url}}{#1}%
    }%
  }{%
    \href{http://dx.doi.org/\thefield{doi}}{#1}%
  }%
}

\DeclareFieldFormat{title}{\usebibmacro{string+doiurlisbn}{\mkbibemph{#1}}}
\DeclareFieldFormat[article,book,inbook,thesis,incollection,inproceedings,misc,online,patent,phdthesis,techreport,unpublished,report]{title}%
    {\usebibmacro{string+doiurlisbn}{\mkbibquote{#1}}}

\definecolor{linkcolor}{rgb}{0.33, 0.41, 0.6}
\usepackage{hyperref}
\hypersetup{
  pdfdisplaydoctitle,
  colorlinks, 
  urlcolor=linkcolor, 
  citecolor=linkcolor,
  linkcolor=linkcolor
}

\newcommand{\red}[1]{{\color{red}#1}}
\definecolor{mygreen}{rgb}{0.0, 0.9, 0.0}
\newcommand{\green}[1]{{\color{mygreen}#1}}

\usepackage{csquotes}

\begin{document}

\title{Does disaggregated electricity feedback
  reduce domestic electricity consumption?
  A systematic review of the literature}

\date{14\textsuperscript{th} March 2016}
\author{Jack Kelly and William Knottenbelt
  \\Department of Computing, Imperial College London, UK.
  Email: jack.kelly@imperial.ac.uk}
\maketitle

\begin{abstract}
  We examine twelve studies on the efficacy of disaggregated energy
  feedback. The average electricity reduction across these studies is
  4.5\%.  However, 4.5\% may be a positively-biased estimate of the
  savings achievable across the entire population because all twelve
  studies are likely to be prone to `opt-in' bias hence none test the
  effect of disaggregated feedback on the \textit{general population}.
  Disaggregation may not be \textit{required} to achieve these
  savings: Aggregate feedback alone drives 3\% reductions; and the
  four studies which directly compared aggregate feedback against
  disaggregated feedback found that \textit{aggregate} feedback is
  \textit{at least as effective} as disaggregated feedback, possibly
  because web apps are viewed less often than in-home-displays (in the
  short-term, at least) and because some users do not trust
  fine-grained disaggregation (although this may be an issue with the
  specific user interface studied).  Disaggregated electricity
  feedback may help a motivated sub-group of the population, which we
  call `energy enthusiasts', to save more energy but fine-grained
  disaggregation may not be necessary to achieve these energy savings.
  Disaggregation has many uses beyond those discussed in this paper
  but, on the specific question of promoting energy reduction in the
  general population, there is no robust evidence that current forms
  of disaggregated energy feedback are more effective than aggregate
  energy feedback.  The effectiveness of disaggregated feedback may
  increase if the general population become more energy-conscious
  (e.g. if energy prices rise or concern about climate change
  deepens); or if users' trust in fine-grained disaggregation
  improves; or if innovative new approaches or alternative
  disaggregation strategies (e.g. disaggregating by \textit{behaviour}
  rather than by \textit{appliance}) out-perform existing feedback.
  This paper also discusses opportunities for new research into the
  effectiveness of disaggregated feedback.
\end{abstract}

\section{Introduction}

Electricity disaggregation estimates the energy consumption of
individual appliances (or load types or behaviours) using data from a
single meter.  One use-case is to estimate an itemised electricity
bill from a single smart meter measuring the whole building's
electricity demand.

Research into electricity disaggregation algorithms began over thirty
years ago~\cite{hart1984, hart1992}.  Today, there is a lot of
excitement about energy disaggregation.  Since 2010 there has been a
dramatic increase in the number of papers published on energy
disaggregation algorithms and since 2013 there have been over 100
papers published each year~\parencite{parson2015overview}.
Disaggregation is big business: In November 2015 disaggregation
provider \href{http://www.bidgely.com}{Bidgely} raised \$16.6 million
USD~\parencite{richardson2015}.  There are now at least 30 companies
who offer disaggregation products and
services~\cite{parson2016industry, kelly2016thesis}.

This paper discusses four main questions:
1) Can disaggregated energy data help an already-motivated
  sub-group of the general population \mbox{(`energy enthusiasts')} to save
  energy?
2) How much energy would the \textit{general population} save if
  given disaggregated data?
3) Is \textit{fine-grained} disaggregation required?
4) For the general population, does disaggregated energy feedback
  enable greater savings than \textit{aggregate} data?

\subsection{An introduction to systematic reviews}

This paper is, to the best of our knowledge, the first systematic
review on the effectiveness of domestic, disaggregated electricity
feedback.

Systematic reviews are common in fields such as medicine and the
social sciences.  Systematic reviews aim to find results which are
robust across \textit{multiple} studies as well as opportunities for
future research.  The process starts with a search, using predefined
criteria, for existing papers.  Results and possible biases are
extracted from each paper, collated and combined.  See Garg et
al.\ \parencite{garg2008} for a discussion of systematic reviews.

There is a distinction between \textit{narrative} reviews and
\textit{systematic} reviews.  Most review articles are narrative
reviews.  These are written by domain experts and contain a discussion
of existing papers.  Narrative reviews are often very valuable.  But
they are rarely explicit about how papers were selected and rarely
attempt a quantitative synthesis of the results.

Systematic reviews aim to cover \textit{all} papers which match
defined criteria relevant to a specific research question.  Systematic
reviews are explicit about how papers were selected and present a
quantitative summary of each paper and a quantitative synthesis of the
results.  Systematic reviews may contain a `meta-analysis' where
results from each study are combined into a single statistical
analysis which provides greater statistical power than any individual
study can deliver.

Systematic reviews are not perfect, of course.  Bias can still creep
in via the selection process; and different statistical analyses may
present different results.

Why bother with systematic reviews? \textit{Replication} is a
essential to the scientific process. Peer review is necessary
\textit{but not sufficient} to ensure that individual studies present
an accurate estimate of the `true' state of the world.  Reviewers rarely, if ever,
attempt to replicate results; possibly because there is insufficient
reward to motivate reviewers to spend time on replication.  Instead it
is left to the community to attempt to replicate results.  Recent
large-scale replication projects suggest that replicable results may
be the \textit{exception} rather than the rule.  An attempt to
replicate results from 98 psychology papers could only replicate 39\%
of the results~\cite{baker2015reproducibility, 2015estimating}.  A
similar study in cancer biology found that only 6 of the results in 53
high-profile papers could be
replicated~\cite{baker2015reproducibility, begley2012drug}.  Hence it
is advisable to exercise appropriate scientific scepticism when
reading any single study; and it is beneficial to collect all papers
on a specific question to identify results which are robust
across studies.

\subsection{Methodology}

Broadly, this paper discusses whether deployment of disaggregation
across the entire population is likely to reduce energy consumption.
We assume that disaggregated data for a population-wide
deployment would be delivered via websites, smart-phone applications
or paper bills.

We found twelve groups of studies on the question of whether
disaggregated energy data helps users to reduce their energy
demand.  These studies are summarised in
Table~\ref{tab:effectiveness-disag}.

We aimed to do an exhaustive search of the literature although it is
not possible to rule out the possibility that we missed studies.  We
used three search engines: Google Scholar, the ACM Digital Library and
IEEE Xplore.  The search terms we used were `disaggregated
[energy|electricity] feedback' and `N[I|A|IA]LM feedback'.  These
searches produced a huge number of results, many of which were not
relevant to our research question.  We manually selected papers which
test the effectiveness of disaggregated electricity feedback.  We
accepted experiments conducted either in a laboratory environment or
in a field test.  We also searched the bibliography sections of papers
to find more papers.  For example, a review article by
Ehrhardt-Mertinez et al.\ \cite{ACEEE2010} contained references to
five relevant studies on disaggregated energy feedback.

\section{Can disaggregated electricity feedback enable \mbox{`energy enthusiasts'} to
  save energy?}

The mean reduction in electricity consumption across the twelve
studies (weighted by the number of participants in each study) is
4.5\%. However, as we will discuss below, this figure is likely to be
positively-biased and has a substantial (although
unquantifiable) amount of uncertainty associated with it.

Aggregating the results by taking the \textit{mean} of the energy
savings across the twelve studies is a crude approach.  It would have
been preferable to do a full meta-analysis where biases are identified
and compensated for \cite{garg2008}.  Davies et al.\
\cite{davis2013setting} did such a meta-analysis for studies on
\textit{aggregate} energy feedback.  But the studies on disaggregated
feedback appear to us to be too varied and, perhaps most
fundamentally, six of the twelve studies only provided a point
estimate of the effect size.  At the very least, a meta-analysis
requires that each study provides a point estimate \textit{and} a
measure of the \textit{spread} of the results.

We must also be explicit about the likely biases in each study.
Please note that this is \textit{not} an attack on the papers in
question!  We appreciate that it is not possible to conduct a `perfect' study.
The real world is messy and researchers cannot control for everything!
Being explicit about the biases allows us to assess how
much trust we should put into the assertion that disaggregated energy
feedback reduces consumption by 4.5\%.

There are several sources of positive bias present in the papers. All
twelve studies are prone to `opt-in' bias, where subjects
self-selected to some extent and so are likely to be more interested
in energy than the general population.

Eight studies did not control for the Hawthorne effect.  This strange
effect is where participants reduce their energy consumption simply
because they know they are in an energy study.  For example, Schwartz
et al.\ \cite{schwartz2013} conducted a controlled study on 6,350
participants, split equally between control and treatment
groups. Subjects in the treatment group received a weekly postcard
saying: \textit{`You have been selected to be part of a one-month
  study of how much electricity you use in your home... No action is
  needed on your part. We will send you a weekly reminder postcard
  about the study...'}  Participants who received these postcards
reduced their consumption by 2.7\%.  Hence studies on disaggregated
energy feedback which do not control for the Hawthorne effect are
likely to over-estimate energy savings attributable to the
disaggregated energy feedback.

Six studies used feedback displays which were probably more
attention-grabbing than the feedback mediums that would be used in a
population-wide roll-out of disaggregated energy feedback. Some
studies gave home-visits to some participants to enable additional
reductions (e.g. \cite{schwartz2014, brown2014HEA,
  HEA2015EUMV-final-report}).  All but two studies were too short to
observe whether energy reductions persist long-term.  Perhaps some
authors experimented with multiple statistical techniques until one
delivered a significant result.  And, finally, eight studies used
sub-metered data, hence avoiding any mistrust of disaggregated
estimates~\cite{churchwell2014}.

As well as being explicit about biases in each study, we must
acknowledge that the \textit{literature} as a whole may be prone to
publication bias.  How many \textit{negative} results exist
unpublished?  Perhaps academics fear that reviewers would reject a
null result? Might companies fear that customers or shareholders would
be driven away?  A study on publication bias in the social sciences
found that positive results are 60\% more likely to be written up than
null results and 40\% more likely to be published~\cite{franco2014}.
They propose that science would benefit from mechanisms to reduce the
effect of publication bias, such as pre-registering experiments.


Despite these sources of bias, there is evidence that energy
disaggregation \textit{can} enable energy savings for `energy
enthusiasts'.  Two large studies illustrate this assertion:

One group of studies analysed the disaggregation service provided by
\href{http://corp.hea.com}{Home Energy Analytics}
(HEA)~\parencite{schmidt2012reductions,
  HEA2012High-Energy-Homes-Summary, HEA2013EUMV-phase-1, brown2014HEA,
  HEA2015EUMV-final-report}. All participants \textit{opted into}
HEA's system and hence could loosely be considered `energy
enthusiasts'.  In total the HEA papers examine 1,623 users.  1,239
used the system for up to 44 months; the rest used the system for one
year.  The average reduction in electricity consumption across all
1,623 `energy enthusiasts' was 6.1\%.  The top-quartile (310
`super-enthusiasts') reduced their electricity consumption by 14.5\%.
But note that none of the HEA studies had a control group.

Another large study was performed in 2014 over three months on 1,685
PG\&E users~\parencite{churchwell2014, bidgely2015}.  Half received an
in-home-display (IHD) and half received access to Bidgely's website
(which includes disaggregation). No statistically significant
reduction in consumption was found across all 1,685 users, despite
positive biases (e.g. users could \textit{choose} between the IHD or
Bidgely).  However, a sub-group of users on a time-of-use (TOU) tariff
(`energy enthusiasts') saved 7.7\%. The TOU group consisted of 142 IHD
users and 136 Bidgely users.

\section{How much energy would the \textit{WHOLE} population save
  if given disaggregated data?}

\input{disagfeedback-table}

All twelve studies suffer from opt-in bias to some extent. Seven
studies have a \textit{high} risk of opt-in bias because participants
sought out the intervention.  As such, the study participants are
unlikely to be representative of the general population.  No `perfect'
correction for opt-in bias exists.

What will the \textit{average} energy saving be across the population
if, say, the majority of the population completely ignores
disaggregated energy feedback but a small sub-population of `energy
enthusiasts' save 4.5\%?

How can we estimate the proportion of `energy enthusiasts' in the
population?  Three studies reported the number of people
\textit{approached} to participate versus the number who
\textit{agreed} to participate~\parencite{wood2003,
  HEA2015EUMV-final-report, sokoloski2015}.  This `opt-in rate' is a
crude estimate on the \textit{lower bound} of the proportion of the
population who are `energy enthusiasts' (because, in order to agree to
participate in an energy study, people probably need to be energy
enthusiasts \textit{and} also have time to participate in the study
\textit{and} be willing to let experimenters into their homes etc.).
The average opt-in rate is 16\%.  This is consistent with
\cite{muragh2014} 
who estimate that 20\% of the population are `[energy]
monitor enthusiasts'.  If 16\% of the population reduced their energy
consumption by 4.5\% then the mean reduction would be 0.7\%.

This may seem rather pessimistic.  We assumed that 84\% of the
population (the `disinterested') would save \textit{no} energy.
Perhaps this is a little unrealistic: We might hope that some
proportion of the `disinterested' group would save a little energy.
Furthermore, we used a crude method to determine a lower bound on the
proportion of `energy enthusiasts' in the population.  But
remember that we have multiple reasons for believing that a 4.5\%
saving across the `energy enthusiast' population is an
\textit{over-estimate}.  We assume that these negative and positive
biases cancel out, although we cannot be sure.

Also note that we simply have \textit{no} good evidence for how the
\textit{general population} would react to disaggregated energy data.
However, related studies have found that effect sizes reported on
opt-in groups are often substantially diminished when studied on the
general population~\parencite{davis2013setting}.


Can we compare these figures to other research?  A
study involving 2,000 Swedish households found that participants who
visited a website which provided user-friendly analysis of their
aggregated electricity consumption reduced their electricity
consumption by 15\% on average~\cite{vassileva2012impact}. The savings
sustained for the duration of the four-year study.  But only 32\% of
those with access to the website visited the website.  Households who
did not visit the website did not reduce their energy
consumption. Hence the average energy reduction across all households
with access to the website was $32\% \times 15\% \approx 5\%$.

\section{Is `fine-grained' disaggregation necessary?}

Much research into disaggregation aims to deliver `fine-grained'
estimated power demand for each appliance at relatively high temporal
resolution (e.g. 0.1~Hz).  Fine-grained disaggregation is complex to
engineer and often computationally expensive to run.  Is it worth the
effort?  \href{http://corp.hea.com}{Home Energy Analytics (HEA)} do
`coarse-grained' disaggregation: they disaggregate energy usage into
five broad categories at \textit{monthly} temporal resolution.
Despite the coarse granularity of the feedback, HEA achieved
significant average reductions in electricity usage of
6.1\%~\parencite{schmidt2012reductions,
  HEA2012High-Energy-Homes-Summary, HEA2013EUMV-phase-1, brown2014HEA,
  HEA2015EUMV-final-report}.  HEA's results tell us that fine-grained
feedback is certainly not \textit{required}.  Fine-grained feedback
enables many use-cases not discussed here but, on the question of the
efficacy of feedback to drive energy reductions, we simply do not know
if fine-grained feedback is more effective because no studies compared
fine-grained against coarse-grained feedback.  Fine-grained feedback
might be \textit{less} effective because some users do not trust it
\cite{churchwell2014}.

\section{Does aggregate or disaggregated
  feedback enable greater savings for the whole population?}

Four studies directly compared aggregate feedback against
disaggregated feedback.  Three of these studies found aggregate
feedback to be \textit{more} effective than disaggregated feedback~\cite{krishnamurti2013IHDs,
  churchwell2014, sokoloski2015}.  The fourth study found
disaggregated feedback and aggregate feedback to be equally
effective~\cite{mccalley2002energy}.  Are there any explanations for
this counter-intuitive result?

Two of the four studies \cite{mccalley2002energy,
  krishnamurti2013IHDs} were synthetic computer simulations and so may
not generalise.

The other two studies were well controlled field
studies~\cite{churchwell2014, sokoloski2015}.  In both field studies,
aggregate feedback was displayed on an always-on IHD whilst
disaggregated data was displayed on Bidgely's website (which has since
been redesigned~\cite{bidgely2015}).  Participants in the
disaggregation groups did not have an IHD.
Sokoloski~\cite{sokoloski2015} found that, on average, participants in
the IHD condition viewed the IHD eight times per day whilst
participants in the disaggregation condition viewed the website only
once per day.  Churchwell et al. \cite{churchwell2014} found a similar
pattern and also reported that some participants did not trust the
fine-grained disaggregated data.  Perhaps aggregate data is not
\textit{intrinsically} more effective than disaggregated data;
instead, perhaps \textit{IHDs} are more effective than
\textit{websites} or mobile apps.

Perhaps \textit{dedicated} displays for disaggregated data may help
enhance efficacy, although this adds costs.  Or, as
Sokoloski~\cite{sokoloski2015} suggests, efficacy may be increased by
\textit{combining} disaggregated feedback presented on a website with
aggregate feedback presented on an IHD.

Furthermore, a meta-analysis of the efficacy of \textit{aggregate}
energy feedback suggests it alone achieves 3\% energy
savings~\cite{davis2013setting}.  This analysis adjusted for
several (but not all) biases.

\section{Suggestions for future research}

There are several gaps in the existing literature.  Below is a list of
potential experiments (more ideas are listed in~\cite{kelly2016thesis}).

No existing field studies compared aggregate feedback against
disaggregated feedback \textit{on the same type of display}.  The
studies which \textit{did} compare aggregate feedback against
disaggregated feedback used an IHD for aggregate feedback and a
website for disaggregated feedback and found that the aggregate
feedback was more effective at reducing energy demand. But we cannot
rule out that this result is simply because users viewed the IHD more
frequently than they viewed the website. Hence it would be valuable to
run an experiment where both the `aggregate' and `disaggregated'
groups received feedback on the same device (e.g. an IHD with a
dot-matrix display to display disaggregated feedback).

A related study would explore the effectiveness of aggregate feedback
presented on an IHD \textit{combined with} access to disaggregated
data on a website; compared to just the IHD. The IHD might pique users'
interest and motivate them to explore their disaggregated energy usage
on a website or smart phone.

Another study would compare fine-grained disaggregated
feedback against coarse-grained disaggregated feedback.

Below is a list of suggestions for how to make future papers on
feedback as useful as possible: If possible, conduct a
\textit{randomised controlled trial}.  Publish as much information as
possible.  How were subjects recruited?  Were subjects selected from
the \textit{general population}?  Did any subjects withdraw during the study
period?  Was there a control group? Did the study control for the
Hawthorne effect and weather?  What sources of bias may influence the
result?  How exactly was feedback presented; and how rapidly did the
information update?  How often did participants view the display?  Was
disaggregated data available from the very beginning of the experiment
or did the disaggregation platform take time to adapt to each home?
Publish the results of \textit{all} the valid statistical analyses
performed; not just the `best' result.  Crucially, please publish some
measure of the \textit{spread} of the result (e.g. the standard
deviation).  Ideally, publish online, full, anonymised results so
researchers can collate your results into a meta-analysis.

\vfill

\section{Conclusions}

Disaggregation has \textit{many} use-cases beyond feedback. This paper
specifically considers a single use-case of disaggregation: Reducing
energy consumption via feedback.  Averaged across the population, there is
evidence that disaggregated feedback \textit{may} help to reduce
electricity consumption by \textasciitilde0.7-4.5\%.  But disaggregation might
not be \textit{necessary} to achieve this saving because aggregate
feedback may be equally effective.  Amongst `energy enthusiasts',
disaggregated feedback might save more energy but fine-grained
disaggregation may not be necessary.

We must emphasise that all we can do is report the current state of
the research. We \textit{cannot} rule out the possibility that
disaggregated feedback is, in fact, more effective than aggregate
feedback.  Neither can we rule out that fine-grained feedback is more
effective than coarse-grained. All we can say is that current evidence
contradicts the first hypothesis and that there is no evidence
available to address the second hypothesis.

Importantly, note that the existing evidence-base is
heterogeneous and has many gaps. Perhaps a large, well controlled,
long-duration, randomised, international study will find that
disaggregated feedback is more effective than aggregate.  

Perhaps users will become more interested in disaggregated data if
energy prices increase or if concern about climate change deepens. Or
perhaps users in fuel poverty will be more likely to act on
disaggregated feedback in order to save money.  Or perhaps users will
trust disaggregation estimates more if accuracy improves or if
designers find ways to communicate uncertain disaggregation estimates.
Or perhaps real-time feedback or better recommendations will improve
performance.  Or perhaps disaggregating by behaviour rather than by
appliance will make disaggregated feedback more effective.

\printbibliography 

\end{document}


%% file: disagfeedback-table.tex
\ifdefined\WORKSHOP
  \newcommand{\fnsize}{\footnotesize}
  \newcommand{\colAwidth}{45mm}
\else
  \newcommand{\fnsize}{\scriptsize}
  \newcommand{\colAwidth}{38mm}
\fi

\newcommand{\rotA}[1]{%
  \multicolumn{1}{c} 
  {\footnotesize
    \rot
      {\makecell[{{p{
              \ifdefined\WORKSHOP 45mm \else 51mm \fi
            }}}]{#1}}}
}

\newcommand{\colA}[1]{
  \multicolumn
    {1}
    {>{\columncolor[gray]{0.9}}C{\colAwidth}}
    {#1}
}

\newcommand{\ttmark}[1]{\ifdefined\WORKSHOP \else \tmark[#1] \fi}

\newcommand{\newRow}{\NN\NN[-2.7mm]}

\ifdefined\WORKSHOP
  \renewcommand{\dbltextfloatsep}{0pt}
\fi

\begin{spacing}{1}
\ctable[
  center,
  star, 
  caption = {Studies on the effectiveness of disaggregated energy feedback.},
  label = tab:effectiveness-disag,
  doinside = \fnsize \setlength{\tabcolsep}{0.5pt} \setlength{\extrarowheight}{0.4mm}
    \renewcommand{\arraystretch}{1}
]{
  C{\colAwidth} 
  >{\columncolor[gray]{0.95}}C{\ifdefined\WORKSHOP 20mm \else 14mm \fi}
  >{\columncolor[gray]{0.9}}C{8mm}
  >{\columncolor[gray]{0.95}}C{8mm}
  >{\columncolor[gray]{0.9}}C{7mm}
  >{\columncolor[gray]{0.95}}C{8mm}
  >{\columncolor[gray]{0.9}}C{8mm}
  >{\columncolor[gray]{0.95}}C{6mm}
  >{\columncolor[gray]{0.9}}C{11mm}
  >{\columncolor[gray]{0.95}}C{8mm}
  >{\columncolor[gray]{0.9}}C{9mm}
  >{\columncolor[gray]{0.95}}C{8mm}
  >{\columncolor[gray]{0.9}}C{6mm}
  >{\columncolor[gray]{0.95}}C{6mm}
  >{\columncolor[gray]{0.9}}C{6mm}
  >{\columncolor[gray]{0.95}}C{6mm}
  >{\columncolor[gray]{0.9}}C{6mm}
}{
  \\[\ifdefined\WORKSHOP -0.5em \else -1em \fi] 
  \tnote[]{\fnsize A dash `-' in a cell means `not applicable
    (NA)' and `?' means `not specified in paper'.}
  \tnote[U]{\fnsize Absolute reductions minus
    reductions for the no-contact control
    (or the most similar group to a no-contact control available).}
  \tnote[R]{\fnsize Recommendations can be `P' for `personalised' or `G' for
    `general' or `\xmark'\ for none given.}
  \tnote[V]{\fnsize Volunteer bias can be `H' for `high' (subjects sought out
    the intervention) or `L' for `low' (subjects were approached by
    the experimenters but only a fraction agreed to participate).}
  \tnote[T]{\fnsize H=hourly, D=daily, M=monthly, Y=yearly,
    B=current billing cycle.}
  \tnote[\#]{\fnsize Paper is silent on this question.  Assume the
    worst.}
  \ifdefined\WORKSHOP
  \else
    \tnote[a]{\fnsize \cite{dobson1992} do not state exactly how households were
      recruited.  They write ``\textit{100 all-electric
        households were qualified from a random sample drawn
        from a population of approximately 8800 such houses}''.  I assume
      households were not \textit{forced} to participate so they must have
      self-selected to some extent.} 
    \tnote[b]{\fnsize A washing machine control was simulated on a computer. The
      reported energy reduction is \textit{only} for the simulated washer.
      The no-feedback-no-goal condition and the
      feedback-no-goal conditions achieved the same reduction (11\%),
      hence the difference in energy savings between those two conditions is 0\%.}
    \tnote[c]{\fnsize One group 
      received both real-time energy feedback for the cooker and a printed information
      pack of general recommendations but this group achieved lower energy savings (8.9\%) than the group
      which only received energy feedback.} 
    \tnote[d]{\fnsize ECOIS-I started with 9 houses but one house was excluded
      because it had solar PV installed.} 
    \tnote[e]{\fnsize \cite{ueno2006ECOIS-I} report that the ``\textit{average ambient
      temperatures before and after installation were 6.4 and
      6.8~\celsius, respectively.  Generally, the power consumption of
      the whole household increases with the fall in
      ambient temperature in winter; hence, it is thought that
      the true effect was more than this 9\% value.}''}
     \tnote[f]{\fnsize Aggregate data was displayed real-time.
       Disaggregated data was not real-time.}
  \fi
\ifdefined\WORKSHOP \\[-0.5em] \bottomrule \fi
}{\FL
\rotA{Study} 
& \rotA{Feedback presentation}
& \rotA{Num. houses in disag. group}
& \rotA{Num. houses in study}
& \rotA{Num. disaggregation categories}
& \rotA{Duration (months of disag)}
& \rotA{Reduction in electricity use\tmark[U] (\%)}
& \rotA{Reduction is for whole house?}
& \rotA{Sample period of meter}
& \rotA{Feedback delay} 
& \rotA{Timing: Historic or Concurrent?}
& \rotA{Time frames for historic\tmark[T]}
& \rotA{Recommendations given?\tmark[R]}
& \rotA{Control group?}
& \rotA{Controlled for Hawthorne?}
& \rotA{Volunteer bias?\tmark[V]}
& \rotA{Controlled for weather?}
\newRow
\colA{
  \ifdefined\WORKSHOP
    ``RECS'' \cite{dobson1992} 
  \else
    ``RECS'' \newline \cite{dobson1992} 
  \fi
}
& dedicated computer 
& \dmake{2}{0}{25} 
& \dmake{4}{0}{100} 
& $\sim 8$ 
& 2 
& \dmake{2}{1}{12.9} 
& \cmark 
& 0.6 sec
& 0 
& H\&C
& HDM 
& \xmark 
& \green{\cmark} 
& \green{\cmark} 
& \green{L}\ttmark{a} 
& \green{\cmark} 
\newRow
\colA{
  \ifdefined\WORKSHOP
    McCalley \& Midden 2002
  \fi
\cite{mccalley2002energy}
} 
& \ifdefined\WORKSHOP 
    Virt. wash. machine 
  \else 
    Virtual washing machine 
  \fi 
\ttmark{b} 
& \dmake{2}{0}{25} 
& \dmake{4}{0}{100} 
& 1 
& - 
& \dmake{2}{1}{0.0} 
& \xmark 
& - 
& 0 
& H\&C 
& - 
& G 
& \green{\cmark} 
& \green{\cmark} 
& \green{L} 
& - 
\newRow
\colA{
  \ifdefined\WORKSHOP
    Wood \& Newborough '03 \cite{wood2003}; \newline
    Mansouri \& Newborough '99 \cite{mansouri1999}
  \else
    \citeauthor{wood2003} 03;
    \citeauthor{mansouri1999} 99
  \fi
} 
& LCD by cooker 
& \dmake{2}{0}{10} 
& \dmake{4}{0}{44} 
& 1 
& $\ge2$ 
& \dmake{2}{1}{12.2} 
& \xmark 
& 15 sec
& 0 
& C 
& - 
& \xmark\ttmark{c} 
& \green{\cmark} 
& \green{\cmark} 
& \green{L} 
& \green{\cmark} 
\newRow
\colA{
``ECOIS-I'' \newline
  \ifdefined\WORKSHOP
    \cite{ueno2006ECOIS-I, ueno2006ACEEE}
  \else
    \cite{ueno2006ECOIS-I}; \newline 
    \cite{ueno2006ACEEE}
  \fi
} 
& Dedicated laptop 
& \dmake{2}{0}{8} 
& \dmake{4}{0}{8}\ttmark{d} 
& 16 
& 2 
& \dmake{2}{1}{9} 
& \cmark 
& 30~min 
& next day 
& H 
& D, 10D
& P 
& \red{\xmark} 
& \red{\xmark} 
& \red{H}\tmark[\#] 
& \green{\cmark}\ttmark{e} 
\newRow
\colA{``ECOIS-II'' \newline
  \ifdefined\WORKSHOP
    \cite{ueno2005ECEEE, ueno2006ECOIS-II, ueno2006ACEEE}
  \else
    \cite{ueno2005ECEEE}; \newline 
    \cite{ueno2006ECOIS-II}; \newline
    \cite{ueno2006ACEEE}
  \fi
} 
& Dedicated laptop 
& \dmake{2}{0}{10} 
& \dmake{4}{0}{19} 
& 16 
& 3 
& \dmake{2}{1}{18} 
& \cmark 
& 30~min
& next day 
& H 
& D, 10D
& P 
& \green{\cmark} 
& \green{\cmark} 
& \red{H}\tmark[\#] 
& \green{\cmark} 
\newRow
\colA{``EnergyLife'' trial 1 \newline
  \ifdefined\WORKSHOP
    \cite{jacucci2009designing, spagnolli2011, gamberini2011}
  \else
    \cite{jacucci2009designing}; \newline
    \cite{spagnolli2011}; \newline
    \cite{gamberini2011}
  \fi
}
& iPhone 
& \dmake{2}{0}{13} 
& \dmake{4}{0}{13} 
& 7 
& 3 
& \dmake{2}{1}{5} 
& \cmark 
& ? 
& 1-2 min 
& H\&C 
& D 
& P 
& \red{\xmark}\tmark[\#] 
& \red{\xmark}\tmark[\#] 
& \red{H}\tmark[\#] 
& \red{\xmark}\tmark[\#] 
\newRow
\colA{``EnergyLife'' trial 2 \newline \cite{gamberini2012}} 
& iPhone 
& \dmake{2}{0}{4} 
& \dmake{4}{0}{4} 
& 7 
& 4 
& \dmake{2}{1}{38} 
& \xmark 
& ? 
& 1-2 min 
& H\&C 
& D 
& P 
& \red{\xmark} 
& \red{\xmark}\tmark[\#] 
& \red{H}\tmark[\#] 
& \red{\xmark}\tmark[\#] 
\newRow
\colA{\href{http://corp.hea.com/results}{Home Energy Analytics}  \newline
  \ifdefined\WORKSHOP
    \cite{HEA2012High-Energy-Homes-Summary, HEA2013EUMV-phase-1,
      brown2014HEA, HEA2015EUMV-final-report}
  \else
    \cite{HEA2012High-Energy-Homes-Summary, HEA2013EUMV-phase-1};
    \newline
    \cite{brown2014HEA, HEA2015EUMV-final-report}
  \fi
} 
& Web \& email \& home visits
& \dmake{4}{0}{1623} 
& \dmake{4}{0}{1623} 
& 5 
& $\le44$ 
& \dmake{2}{1}{6.1} 
& \cmark 
& hourly 
& 0 
& H 
& Y 
& P 
& \red{\xmark} 
& \red{\xmark} 
& \green{L} 
& \green{\cmark} 
\newRow
\colA{\href{http://www.bidgely.com}{Bidgely} 
  \ifdefined\WORKSHOP
    2013 \newline
    \cite{chakravarty2013, gupta2014}
  \else
    \newline
    \cite{chakravarty2013}; \newline \cite{gupta2014}
  \fi
} 
& Web, mobile, email 
& \dmake{2}{0}{163} 
& \dmake{4}{0}{328} 
& $\ge3$? 
& - 
& \dmake{2}{1}{6} 
& \cmark 
& 30 sec \newline \& 1 hr 
& 0\ttmark{f} 
& H\&C\ttmark{f} 
& DBY 
& P 
& \green{\cmark} 
& \red{\xmark} 
& \red{H} 
& \green{\cmark} 
\newRow
\colA{PG\&E Pilot
  \ifdefined\WORKSHOP
    2014 \newline
    \cite{churchwell2014, bidgely2015}
  \else
    \newline \cite{churchwell2014}; \newline \cite{bidgely2015}
  \fi
} 
& Web, mobile, email 
& 844 
& 1685 
& $\ge3$? 
& 3 
& \dmake{2}{1}{2.1} 
& \cmark 
& 30 sec 
& 0\ttmark{f} 
& H\&C\ttmark{f} 
& DBY 
& P 
& \green{\cmark} 
& \red{\xmark} 
& \red{H} 
& \green{\cmark}
\newRow
\colA{
  \ifdefined\WORKSHOP
    Schwartz et al. 2014
  \fi
\cite{schwartz2014}
} 
& Web, mob, TV
& \dmake{2}{0}{6} 
& \dmake{4}{0}{6} 
& $\sim 10$ 
& 18 
& \dmake{2}{1}{7.8} 
& \cmark 
& ? 
& 0? 
& H\&C 
& ? 
& ? 
& \red{\xmark} 
& \red{\xmark} 
& \red{H} 
& \red{\xmark} 
\newRow
\colA{
  \ifdefined\WORKSHOP
    Sokoloski 2015
  \fi
\cite{sokoloski2015}
} 
& Web, mob, email
& \dmake{2}{0}{12} 
& \dmake{4}{0}{70} 
& $\ge3$? 
& 0.75 
& \dmake{2}{1}{3} 
& \cmark 
& 30 sec 
& 0\ttmark{f} 
& H\&C\ttmark{f} 
& DBY
& P 
& \green{\cmark} 
& \red{\xmark} 
& \green{L} 
& \green{\cmark} 
\newRow
%
}
\end{spacing}
